\documentclass[runningheads]{svmult}

\usepackage{makeidx}   
\usepackage{graphicx}  
\usepackage{subeqnar}  
\usepackage{multicol}  
\usepackage{physprbb}  
\makeindex             

\newcommand{\pgla}{\mathrel{\hbox{\rlap{\hbox{\lower4pt\hbox{$\sim$}}}\hbox{$<$}}}}
\newcommand{\pgga}{\mathrel{\hbox{\rlap{\hbox{\lower4pt\hbox{$\sim$}}}\hbox{$>$}}}}

\begin{document}
\title*{Intermediate-age globular clusters in merger remnants}
%
%
%
%
%
\author{Paul Goudfrooij}
%
%
%
\institute{Space Telescope Science Institute, 3700 San Martin Drive,
  Baltimore, MD 21218, USA}

\maketitle              

\begin{abstract}
Recent observations of globular clusters in intermediate-age (2--4 Gyr old)
merger remnants have provided the hitherto ``missing link'' between between
young merger remnants and normal elliptical galaxies. The luminosity functions
(LFs) of the ``blue'' globular clusters exhibit a Gaussian shape centered on
$M_V \sim -7.2$ as found for giant ellipticals, while the ``red'' clusters
exhibit LFs that are more extended towards both the bright and faint ends, as
expected for a cluster population of intermediate age. Specific frequencies
for these systems have been calculated by evolving both the LF of the ``red''
clusters and the parent galaxy luminosity to an age of 10 Gyr. They are found
to be consistent with the values found for giant ellipticals in poor
clusters. These findings support the viability of the ``merger
scenario'' for forming the ``red'' GCs in giant early-type galaxies. 
\end{abstract}

\section{Introduction}
This workshop did a great job in re-emphasizing that the study of
globular clusters (GCs) around galaxies provides relevant and unique
information regarding our understanding of the formation history of
galaxies. Their nature as simple stellar population significantly 
simplifies the determination of their ages and metallicities relative
to that of stellar populations that constitute the integrated light of
their parent galaxies.  

A well-known, interesting feature of the GC systems of giant early-type
galaxies is the presence of bimodal color distributions, providing obvious
evidence for the occurrence of a `second event' during the formation
history of these galaxies. To date, three main competing concepts to
explain this phenomenon have been proposed in the literature. The first
one (and the only one to have actually {\it predicted\/} this bimodality) 
is the ``merger scenario'' by Ashman \& Zepf \cite{ashzep92}, who modeled the
properties of GCs forming in a major merger of gas-rich spirals (together
with the  giant elliptical host galaxies themselves). They figured that the
pre-enriched gas associated with the spiral disks would form `red' (i.e.,
metal-rich) GCs during the merger. 
The two other popular concepts are the ``accretion scenario'' \cite{cote+98} 
and the ``multi-phase collapse scenario'' \cite{forb+97}, 
both of which do not involve new GCs formed in a merger.  More details on
these three scenarios (incl.\ their pros and cons) are discussed in A.\
Kundu's contribution to this volume.   

Given these rather different points of view, it is important to review the
direct evidence presented by observations of merger remnants to test the
predictions of the ``merger scenario'' for forming the {\sf red} GCs in giant
early-type galaxies. This is the main purpose of the remainder of this paper.  

Until just a year or so ago, strong evidence for luminous GCs having formed during
major, gas-rich mergers was limited to a half dozen cases, namely 2 ongoing
mergers (NGC\,4038/4039~\cite{whisch95}, 
NGC~3256~\cite{zepf+99}, 
and 4 {\it young\/} merger remnants (NGC~1275~\cite{brod+98}, 
NGC~3921~\cite{schw+96}, 
NGC 3597~\cite{carl+99}, 
NGC~7252~\cite{whit+93}).
These authors determined ages of the luminous GCs in these galaxies from
photometric and/or spectroscopic data. Their ages are typically $\pgla$\,500
Myr, a time at which most GC destruction and evolution processes have only
just started.  

It is therefore perhaps not surprising that major sources of debate in the
context of the nature of the red GCs in giant ellipticals are:\ {\it (i)\/}
whether or not the GCs formed in mergers will reach old age, {\it (ii)\/}
whether or not the metallicities of GCs formed in mergers are compatible
with those of `red' GCs in old ellipticals, and {\it (iii)\/} whether or
not the luminosity functions and specific frequencies of GCs formed in
mergers will become compatible with those of `red' GCs in old ellipticals
at the latter's age. We discuss these items below along with very recent
results on the GC systems of more evolved merger remnants. 

%
\section{The IMF and Metallicity of GCs Formed in Mergers}
One important issue in figuring out whether luminous GCs can survive to old
age is the shape of the initial mass function (IMF) of the (individual)
GCs. Brodie et al.\ \cite{brod+98} 
obtained spectra of luminous GCs in NGC~1275 and found H$\gamma$ and
H$\delta$ equivalent widths that are somewhat larger (and $B\!-\!R$ colors
that are somewhat bluer) than those of the 1995 version of the Bruzual \&
Charlot population synthesis models in the appropriate age range, both for
Salpeter or Scalo IMFs. They showed that the large Balmer line
equivalent widths could only be brought into agreement with those models by
assuming a truncated IMF, e.g., 
involving a mass range of 2--3 M$_{\odot}$. For such a high low-mass
cutoff, they argue that these clusters will fade away in only $\sim$\,1 
Gyr. On the other hand, Schweizer \& Seitzer~\cite{schsei98} showed that
similarly old ($\sim$\,500 Myr) GCs in NGC~7252 show Balmer line strengths 
that are well reproduced by the Bruzual \& Charlot (1996; BC96) models (which
indeed produce larger Balmer equivalent widths than their 1995 models). Also,  
Gallagher \& Smith~\cite{galsmi99} 
found no need to invoke a truncated IMF to explain the Balmer line
strengths in young clusters in the starburst galaxy M\,82. But then again,
a detailed study of the very luminous star cluster M\,82--F~\cite{smigal01}
revealed that a lower mass cutoff of 2--3 M$_{\odot}$ is
required (for a Salpeter IMF) to match their observations. Obviously, 
the IMF among luminous young star clusters may not be universal. 

Another open question until a year or so ago was whether or not the
metallicities of GCs formed 
in mergers are compatible with those of `red' GCs in similarly massive `old'
ellipticals. Metallicity estimates from spectra of young  
GCs in NGC~1275 and NGC~7252 were hampered by relatively large uncertainties, 
since $\sim$\,500 Myr old populations are dominated by A-type
stars with intrinsically weak metallic features. 

Observations of more evolved merger remnants were obviously necessary to make
substantial progress on these issues.  
However, intermediate-age merger remnants (with ages of 1\,--\,5 Gyr) are more
difficult to identify than younger remnants since the main body of the
former has had enough time to relax into a relatively symmetric
configuration and the newly-formed stellar populations have had time to
fade and redden~\cite{schsei92}. 
%
We~\cite{goud+01a} recently conducted multi-slit
spectroscopy of GC candidates in NGC~1316, an early-type galaxy that is an
obvious merger remnant~\cite{schw80}. We discovered the 
presence of $\sim$\,10 GCs associated with NGC~1316 that have luminosities
higher than that of $\omega$\,Cen (up to $\sim$\,10 times that,
actually). Our measurements of H$\alpha$ and the Ca\,{\sc ii} triplet in
the spectra of the brightest GCs showed them to have {\it solar metallicity\/} (to
within 0.15 dex) and to have an age of 3.0 $\pm$ 0.5 Gyr. These are
obviously GCs that have formed from enriched gas during a major merger. This
reinforces the view that luminous GCs formed during mergers do not
necessarily have high low-mass cutoffs to their IMFs. Another important
implication is that these GCs can actually survive dynamical disruption
processes taking place during and after the merging process (see below). 

\section{The Luminosity Function of GCs Formed in Mergers}
\label{s:LFs}
The luminosity functions (LFs) for GCs in young mergers (such as the
Antennae or NGC~7252) are power laws with indices of $\alpha\sim
-2$~\cite{mill+97}; \cite{whit+99}. This distribution
contrasts markedly with the Gaussian profiles found for LFs of old GC
systems such as that of the Milky Way or giant 
ellipticals~\cite{harr91}.
However, various destruction mechanisms ---such as 2-body evaporation, bulge
and disk shocking, and stellar mass loss--- should remove the less massive
and/or more diffuse clusters as the galaxy ages. Hence, a system of GCs
created during a merger may eventually yield a peaked distribution of
clusters similar to what is seen for old GCs~\cite{falzha01},
\cite{goud+01b}. Intermediate-age remnants, 
such as NGC~1316~\cite{goud+01a} and NGC~3610~\cite{schsei92}, 
offer the possibility of actually observing this process in action.
While the blue, metal-poor GCs in these galaxies should have an LF
similar to that of old GCs in elliptical galaxies (peaking at $M_V$ $\approx$
$-$7.2~\cite{whit97}), the LFs of the red, metal-rich clusters are
expected to extend to both brighter and fainter luminosities. 

\begin{figure}
\begin{center}
\includegraphics[width=0.99\textwidth]{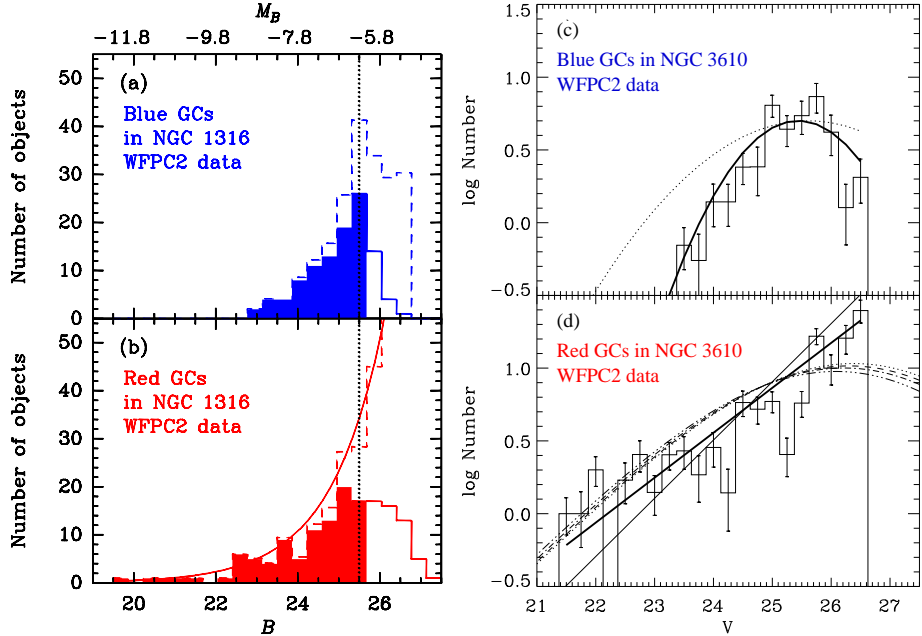}
\end{center}
\caption[]{Luminosity functions (LFs) of GC candidates in NGC~1316 and
NGC~3610 using {\it HST/WFPC2\/} data, taken from Goudfrooij et al.\
\cite{goud+01a} and Whitmore et al.\ \cite{whit+02}, respectively. The top
left panel {\sf 
(a)} shows the LF for the {\it blue\/} GC candidates in NGC~1316
(with $1.00 \leq B\!-\!I \leq 1.55$), panel {\sf (b)} does so for the {\it
red\/} ($1.75 \le B\!-\!I \le 2.50$) GC candidates in NGC~1316, panel
{\sf (c)} does so for the {\it blue\/} GC candidates in NGC~3610, and 
panel {\sf (d)} does so for the {\it red\/} GC candidates in NGC~3610. 
For panels {\sf (a)} and {\sf (b)}, the top axis unit is absolute 
magnitude $M_B$ at the distance of NGC~1316. The histograms are filled for
magnitudes below the 50 per cent completeness limit, and open beyond
it. Dashed lines in panels {\sf (a)} and {\sf (b)} show the effect of
completeness and background corrections, while the dotted line at $B$ =
25.5 depicts the expected LF peak for the old GCs in NGC~1316. The data in
panels {\sf (c)} and {\sf (d)} are corrected for completeness. Solid
(thick) curves in panels {\sf (b)} and {\sf (d)} are power-law fits to the
completeness-corrected luminosity functions. Other curves in panels {\sf
  (c)} and {\sf (d)} are described in ref.\ \cite{whit+02}.} 
\label{f:GCLFs}
\end{figure}

This prediction has recently been verified for two intermediate-age
merger remnants, providing important evidence that mergers {\it
  can\/} indeed create ``red'' GC subsystems that survive destruction. 
Goudfrooij et al.\ \cite{goud+01b} used WFPC2 photometry of the GC
system of the 3-Gyr-old merger remnant NGC~1316. By using appropriate color
cuts in the color-magnitude diagram of NGC~1316's GC system, they showed that
indeed, the LF for the blue GCs is approximately Gaussian centered on the
expected turnover magnitude, while the LF for the red GCs is more nearly
power-law shaped down to the completeness limit of the data (see
Fig.~\ref{f:GCLFs}a,b). Likewise, Whitmore et al.\ \cite{whit+02} performed deep
WFPC2 photometry of the GC system of the candidate 4-Gyr-old merger
remnant NGC~3610 \cite{schsei92}, \cite{whit+97}. They
found a very similar result, with the red GCs still showing a power-law LF
shape (see Fig.~\ref{f:GCLFs}c,d). 

\begin{figure}
\begin{center}
\includegraphics[width=0.6\textwidth]{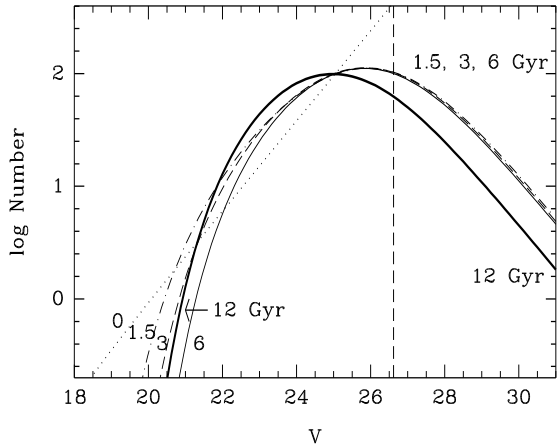}
\end{center}
\caption[]{Predicted evolution of LFs based on a combination
of the Fall \& Zhang \cite{falzha01} cluster disruption models and the BC96
population synthesis models. Solar metallicity is assumed for
the young and intermediate-age GCs while 0.02 solar is assumed
for the 12 Gyr population. The LFs are normalized at $V =
25$. Note that the fading predicted by the BC96
models almost exactly cancels out the increase in the mean mass
predicted by the Fall and Zhang models, resulting in nearly identical
distributions at the faint end for the 1.5, 3, and 6 Gyr models. Figure taken
from ref.\ \cite{whit+02}.} 
\label{f:LFevol}
\end{figure}

An important next step for progress in this field will be finding evidence
for a metamorphosis of the power-law-like LF of those red GCs into a
Gaussian shape as seen in old GC systems. This metamorphosis is now predicted
theoretically (\cite{falzha01}, esp.\ their Fig.\ 4). Quantitative
estimates of the LF shape evolution can be made by combining GC disruption
model predictions (of {\it mass\/} functions) with $M/L$ ratio predictions
from stellar population models. This was done by Whitmore et
al.\ \cite{whit+02} who showed that after 1.5 -- 2 Gyr, the effect of the
shift of the peak of the GC mass function toward smaller masses (due to
disruption processes) to the LF is essentially compensated by the luminosity
fading of GCs.  The net prediction is that at a given metallicity, the peak of
the LF remains nearly constant with time in the range 1.5 -- 12 Gyr. This
effect is
depicted in Fig.~\ref{f:LFevol}. Note that the LF turnover for
metal-rich, intermediate-age GCs is predicted to be $\sim$\,1 mag fainter
than that of the old, metal-poor GCs. This prediction will be tested in a
forthcoming {\it HST\/} program utilizing the enormous discovery power of
the Advanced Camera for Surveys {\it (ACS)}.

\section{Specific Frequencies of GCs formed in Mergers}
One of the arguments often made against the concept that red GCs in giant
ellipticals have been formed in gas-rich mergers is that the 
specific frequencies ($S_N$s) of GCs in merger remnants are low relative to
those in giant ellipticals. I would like to emphasize two things in this
context. First, the number of GCs in the definition of $S_N$ was defined
for ``old'' GC systems with a Gaussian LF, i.e., twice the number of GCs
brighter than the turnover magnitude~\cite{harvdb81}. 
Hence, an appropriate calculation of $S_N$ for merger remnants
requires a separate approach for `blue' and `red' GCs. The `blue' GCs
should be counted the normal way, but the luminosities of the `red' GCs
{\it and of the host galaxy\/} need to be faded to those appropriate for an
age of (say) 10 Gyr.  Also, the `red' GCs need to be counted up to the
turnover magnitude appropriate for a 10 Gyr old population at the observed
GC metallicity, calculated as described in Section~\ref{s:LFs}. Specific
frequencies for the intermediate-age merger remnants NGC~1316,
NGC~1700~\cite{whit+97}, and NGC~3610 calculated this way are listed in 
Table~\ref{t:SNs}. A range of $S_N$ values is given for each galaxy,
indicating the allowed range of the current fraction (by mass) of
intermediate-age stars in the parent galaxy. 

\begin{table}
\caption{Specific Frequencies $S_N$ for Intermediate-Age Merger Remnants}
\begin{center}
\renewcommand{\arraystretch}{1.2}
\setlength\tabcolsep{5pt}
\begin{tabular}{@{}ccl@{}}
\hline\noalign{\smallskip}
Galaxy & $S_N$ & Ref.\ of GC counts\\
\noalign{\smallskip}
\hline
\noalign{\smallskip}
NGC 1316 & 2.1 $\cdots$ 3.4 & \cite{goud+01b} \\
NGC 1700 & 1.6 $\cdots$ 2.8 & \cite{whit+97} \\
NGC 3610 & 1.7 $\cdots$ 2.9 & \cite{whit+02} \\
\hline
\end{tabular}
\end{center}
\label{t:SNs}
\end{table}

Second, the $S_N$ values for these three galaxies should be compared with
those of `normal' ellipticals in similar environment densities. NGC~1316 is in a
poor group behind the Fornax cluster~\cite{drin+01}, \cite{goud+01a}. 
NGC~3610 and NGC~1700 are in poor groups with 5 and
7 members,  respectively~\cite{garc93}. Hence their $S_N$ values should be
compared with the mean $S_N$ of ellipticals in poor groups, which is 2.6 $\pm$
0.5~\cite{harr91}. A glance at Table~\ref{t:SNs} then shows that {\it the
  specific frequencies of GCs in intermediate-age merger remnants are 
consistent with those of `old' ellipticals}.

%


\begin{thebibliography}{26.}
\addcontentsline{toc}{section}{References}
\bibitem{ashzep92} K.M. Ashman, S.E. Zepf: ApJ \textbf{384}, 50 (1992)
\bibitem{cote+98} P. C\^ot\'e, R.O. Marzke, M.J. West: ApJ \textbf{501}, 554
  (1998) 
\bibitem{forb+97} D.A. Forbes, J.P. Brodie, C.J. Grillmair: AJ \textbf{113},
  1652 (1997) 
\bibitem{whisch95} B.C. Whitmore, F. Schweizer: AJ \textbf{109}, 960 (1995)
\bibitem{zepf+99} S.E. Zepf, K.M. Ashman, J. English, et al.: AJ \textbf{118},
  752 (1999) 
\bibitem{brod+98} J.P. Brodie, L.L. Schroder, J.P. Huchra, et al.: AJ
  \textbf{116}, 691 (1998)
\bibitem{schw+96} F. Schweizer, B.W. Miller, B.C. Whitmore, et al.: AJ
  \textbf{112}, 1839 (1996) 
\bibitem{carl+99} M.N. Carlson, J.A. Holtzman, C.J. Grillmair, et al.: AJ
  \textbf{117}, 1700 (1999)
\bibitem{whit+93} B.C. Whitmore, F. Schweizer, C. Leitherer, et al.: AJ
  \textbf{106}, 1354 (1993) 
\bibitem{schsei98} F. Schweizer, P. Seitzer: AJ \textbf{116}, 2009 (1998) 
\bibitem{galsmi99} J.S. Gallagher III, L.J. Smith: MNRAS \textbf{304}, 540
  (1999) 
\bibitem{smigal01} L.J. Smith, J.S. Gallagher III: MNRAS \textbf{326}, 1027 (2001)
\bibitem{schsei92} F. Schweizer, P. Seitzer: AJ \textbf{104}, 1039 (1992) 
\bibitem{goud+01a} P. Goudfrooij, J. Mack, M. Kissler-Patig, et al.: MNRAS
  \textbf{322}, 643 (2001a)
\bibitem{schw80} F. Schweizer: ApJ \textbf{237}, 303 (1980)
\bibitem{mill+97} B.W. Miller, B.C. Whitmore, F. Schweizer, S.M. Fall: AJ
  \textbf{114}, 2381 (1997)
\bibitem{whit+99} B.C. Whitmore, Q. Zhang, C. Leitherer, et al.: AJ
  \textbf{118}, 1551 (1999) 
\bibitem{harr91} W.E. Harris: ARAA \textbf{29}, 543 (1991)
\bibitem{falzha01} S.M. Fall, Q. Zhang: ApJ \textbf{561}, 751 (2001)
\bibitem{goud+01b} P. Goudfrooij, M.V. Alonso, C. Maraston, et al.: MNRAS
  \textbf{328}, 237 (2001b)
\bibitem{whit97} B.C. Whitmore:\ in \emph{The Extragalactic Distance Scale},
ed.\ by M. Livio, M. Donahue \& N. Panagia (STScI, Baltimore), 
p.\ 254
\bibitem{whit+02} B.C. Whitmore, F. Schweizer, A. Kundu, B.W. Miller: AJ
  \textbf{124}, 147 (2002)
\bibitem{whit+97} B.C. Whitmore, B.W. Miller, F. Schweizer, et al.: AJ
  \textbf{114}, 1797 (1997) 
\bibitem{harvdb81} W.E. Harris, S.\ van den Bergh: AJ \textbf{86}, 1627 (1981)
\bibitem{drin+01} M.J. Drinkwater, M.D. Gregg, M. Colless: ApJ  \textbf{548},
  L139 (2001) 
\bibitem{garc93} A.M. Garcia: A\&AS \textbf{100}, 47 (1993) 

\end{thebibliography}
\end{document}